\documentclass[12pt]{article}

\usepackage{amsmath}
\usepackage{amssymb}
\usepackage{amsfonts}
\usepackage{graphicx}
\usepackage{putex}

\numberwithin{equation}{section}

\newcommand{\bC}{\ensuremath{\mathbb{C}}}

\newcommand{\bH}{\ensuremath{\mathbb{H}}}

\newcommand{\bR}{\ensuremath{\mathbb{R}}}

\newcommand{\bZ}{\ensuremath{\mathbb{Z}}}

\newcommand{\scH}{\ensuremath{\mathcal{H}}}

\newcommand{\scN}{\ensuremath{\mathcal{N}}}

\newcommand{\Teichmuller}{Teichm\"{u}ller }

\newcommand{\Li}{\ensuremath{\textrm{Li}}}

\newcommand{\sfX}{\ensuremath{\mathsf{X}}}
\newcommand{\sfY}{\ensuremath{\mathsf{Y}}}
\newcommand{\sfZ}{\ensuremath{\mathsf{Z}}}

\newcommand{\sfx}{\ensuremath{\mathsf{x}}}
\newcommand{\sfy}{\ensuremath{\mathsf{y}}}
\newcommand{\sfz}{\ensuremath{\mathsf{z}}}

\newcommand{\sfL}{\ensuremath{\mathsf{L}}}
\newcommand{\sfR}{\ensuremath{\mathsf{R}}}

\newcommand {\beq}{\begin{eqnarray}}\newcommand {\eeq}{\end{eqnarray}}
\newcommand{\frakm}{\ensuremath{\mathfrak{m}}}
\newcommand{\frakl}{\ensuremath{\mathfrak{l}}}

\def\low#1{\genfrac{}{}{0pt}{}{}{\overbrace{\scriptstyle #1}}}
\def\high#1{\genfrac{}{}{0pt}{}{\underbrace{\scriptstyle #1}}{}}

\def\De{{\Delta}}

\def\Vg{{V_{\rm geom}}}
\def\Vt{{V_{\rm trace}}}

\begin{document}

\preprint{PUPT-11-2383}

\institution{TITECH}{Department of Mathematics, Tokyo Institute of Technology, Tokyo 152-8551, Japan
}

\institution{PCTS}{Princeton Center for Theoretical Science, Princeton University, NJ 08544, USA}

\title{
Semiclassical Analysis of the 3d/3d Relation
}

\authors{Yuji Terashima\worksat{\TITECH} and Masahito Yamazaki\worksat{\PCTS}}

\abstract{
We provide quantitative evidence for our previous conjecture
which states an equivalence of the partition function of a 3d $\scN=2$ 
gauge theory on a duality wall and that of the $SL(2,\bR)$ 
Chern-Simons theory on a mapping torus, for a class of examples 
associated with once-punctured torus.
In particular, 
we demonstrate that a limit of the 3d $\scN=2$ partition function reproduces 
the hyperbolic volume and the Chern-Simons invariant of the mapping torus.
This is shown by analyzing the classical limit
of the trace of an element of the mapping class group 
in the Hilbert space of the quantum \Teichmuller theory.
We also show that the subleading correction to the partition function
reproduces the Reidemeister torsion.
}

\maketitle

%%%%%%%%%%%%%%%%%%%%%%%%%%%%%%%%%%%%%%%%%%%%%%%%%%%%%%%%%%%%%%%%%%%%%%%%
%%%%%%%%%%%%%% new part %%%%%%%%%%%%%%%%%%%%%%%%%%%%%%%%%%%%%%%%%%
%%%%%%%%%%%%%%%%%%%%%%%%%%%%%%%%%%%%%%%%%%%%%%%%%%%%%%%%%%%%%%%%%%%%%%%%

\section{Introduction} \label{sec.intro}

In our previous paper \cite{Terashima:2011qi}, based on previous works
\cite{Drukker:2010jp,Hosomichi:2010vh} we proposed an
equivalence of the partition functions of two 3d theories:
one is a 3d supersymmetric $\scN=2$ theory on a squashed 3-sphere $S^3_b$, where the 3d theory
is realized as a duality 1/2 BPS domain wall inside a 4d $\scN=2$ theory;
another is the 3d bosonic $SL(2,\bR)$ Chern-Simons theory defined on
a mapping torus $M_3$.\footnote{In \cite{Terashima:2011qi}, we considered
two possibilities: $M_3$ is either $\Sigma\times I$ or a mapping torus.
In this paper we focus on the 
latter.} Schematically, our relation is written as
\beq
Z_{\textrm{3d $\scN=2$ theory}}[S^3_b]=Z_{\textrm{3d $SL(2,\bR)$
CS}}[M_3] \, ,
\label{Z=Zintro}
\eeq
where the parameter $b$ is related to the level $k$ of the Chern-Simons
theory.\footnote{More precisely, the right-hand side should be defined by a path
integral over the \Teichmuller component of the moduli space of flat
connections. Such a subtlety, however, does not play a role in the semiclassical
analysis of this paper.}
This relation should arise from the dimensional reduction of the 6d
$(2,0)$ theory on $S^3\times M_3$, and is a 3d/3d counterpart of the
4d/2d correspondence (Alday-Gaiotto-Tachikawa (AGT) conjecture) \cite{Alday:2009aq}.
See \cite{Galakhov:2011hy} for a recent discussion, and
\cite{Dimofte:2010tz,Gadde:2011ik,Nishioka:2011dq} for related proposals.

In \cite{Terashima:2011qi}, we outlined the derivation of the relation 
\eqref{Z=Zintro} using a chain of connections with quantum Liouville and \Teichmuller
theories.
While suggestive, this argument should not be regarded as a proof,
since it relies on several conjectures existing in the literature.
It is therefore highly desirable to provide more quantitative evidence
for this conjecture.

In this paper we perform quantitative checks of our proposal in the
example of the once-punctured torus. 
In particular, we are going to show that the $b\to 0$ limit 
(which is the classical limit $k\to \infty$ of the $SL(2,\bR)$ Chern-Simons theory)
of the partition function of 3d $\scN=2$ theories reproduces
the hyperbolic volume and the Chern-Simons invariant\footnote{This is an
invariant of a hyperbolic 3-manifold, and takes values in real
numbers modulo half-integers.} of the mapping torus:
\beq
Z_{\textrm{3d } \scN=2 \textrm{ theory}}[S^3_b] \to 
  \exp\left[ \frac{1}{2\pi b^2} \left(\textrm{Vol}(M_3)
 + 2\pi^2 i\textrm{CS}(M_3) \right)\right],
\quad \textrm{when  } b\to 0 \,  .
\label{mainclaim}
\eeq
Since the right-hand side is known to be the classical limit of the 
partition function of $SL(2,\bR)$ Chern-Simons theory on $M_3$ \cite{Gukov:2003na}, 
\eqref{mainclaim} is nothing but the $b\to 0$ limit of \eqref{Z=Zintro}.

The rest of the paper is organized as follows. In Sec.\ \ref{sec.idea}
, we summarize in more detail the results of this paper.
In Sec.\ \ref{sec.semiclassical} we study the semiclassical limit of
the right-hand side of \eqref{Z=Zintro} as a trace in quantum 
\Teichmuller theory. The result is then shown to be equivalent to the
geometric potential for hyperbolic volume.
We conclude in Sec.\ \ref{sec.conclusion} with short remarks. 
We also include several appendices on quantum dilogarithm and hyperbolic
geometry.

\section{Summary of Results} \label{sec.idea}

Let us first summarize the main results of the paper in more detail. 

In the case of the once-punctured torus, the corresponding 3d $\scN=2$ 
theories and their partition functions has been described in Sec.\ 4 of 
\cite{Terashima:2011qi}; the gauge theory is a quiver gauge theory where 
the $T[SU(2)]$ theory (and mass deformation thereof) is 
glued together by gauging global symmetries,
with Chern-Simons 
terms added for the corresponding gauge fields.

The important result here, which is first proposed in \cite{Drukker:2010jp},
worked out concretely in \cite{Hosomichi:2010vh} and discussed in 
more generally in \cite{Terashima:2011qi}, is 
that the 3d $\scN=2$ partition function coincides with an
expectation value of an operator in quantum \Teichmuller
theory.\footnote{As explained in \cite{Terashima:2011qi}, this is
statement is not yet completely justified when we need to glue two
theories by gauging global symmetries of the Coulomb branch, which
are quantum symmetries of the theory. \label{footnote1}}
Given a punctured Riemann surface $\Sigma$,
\Teichmuller theory gives an associated Hilbert space 
$\scH_T(\Sigma)$, together with an action of an element $\varphi$ 
of the mapping class group of $\Sigma$.
We can then define a trace of $\varphi$ in the Hilbert space, 
and this coincides with the partition function of the 3d $\scN=2$ theory:
\beq
Z_{\textrm{3d } \scN=2 \textrm{
theory}}[S^3_b]=\textrm{Tr}_{\scH_T(\Sigma)}(\varphi) \, .
\eeq

We are going to analyze the semiclassical limit of our trace 
$\textrm{Tr}(\varphi)$.
The result is given by
\beq
\textrm{Tr}(\varphi) \to \int dx dy du dv \, \exp\left[ 
\frac{1}{2\pi i b^2} V_{\rm trace}(x,y,u,v;h)\right]\, ,
\eeq
where $x, y, u, v$ are a set of parameters and $h$ is a parameter associated with the puncture of $\Sigma$,
and corresponds to a mass parameter in 3d $\scN=2$ theory.
$\Vt$ is quadratic with respect to $u$ and $v$, and 
after extremizing with respect to these variables, we have a
function $\Vt (x,y;h)$:
\beq
\textrm{Tr}(\varphi)\to \int dx dy \, \exp\left[\frac{1}{2\pi i b^2}
\Vt(x,y;h)\right]\, .
\label{Vxy}
\eeq
We will find that this is a linear function with respect to $h$.

Let us now describe the other side of \eqref{Z=Zintro}.
The mapping torus is a 3-manifold defined from $\Sigma$ and $\varphi$.
In particular, when $\varphi$ satisfies 
$\big| \textrm{tr}(\varphi)\big|>2$, $M_3$
admits a finite volume, complete hyperbolic metric \cite{Otal}. 
We can triangulate $M_3$ into ideal 
tetrahedra. The shape of each ideal tetrahedron is specified by a 
parameter called the modulus of the tetrahedron, and 
these parameters should satisfy a set of consistency conditions.
These conditions are generated from the derivatives of a single
potential $\Vg$ \cite{Rivin}. At the extremal point this potential reproduces the
combination $\textrm{Vol}(M_3)+2\pi^2 i \textrm{CS}(M_3)$ of the
3-manifold $M_3$. 

We will demonstrate 
two facts. First, we show (up to constant terms)
\beq
V_{\rm trace}\Big|_{h=0}=V_{\rm geom} \, ,
\eeq
where we have identified the parameters $x, y$ of \eqref{Vxy}
with the moduli of tetrahedra.
This in particular implies \eqref{mainclaim} for $h=0$,
since the hyperbolic volume (plus the Chern-Simons invariant) is given by the critical value of $\Vg$.

Second, for $h\ne 0$ 
we show that 
\beq
V_{\rm trace}(x,y;h)=V_{\rm geom}(x,y;h) \, ,
\eeq
where the potential $\Vg(x,y;h)$ gives a 1-parameter deformation of
hyperbolic structure \cite{NeumannZagier}.
The parameter $h$ is identified with the longitude parameter $\frakl$,
and serves as a Lagrange multiplier\footnote{We would like to thank
T.~Dimofte for suggesting the possibility that the puncture parameter
is the longitude parameter.}. 
Its dual variable, the meridian parameter $\frakm$, is given by
\beq
\frakm=\frac{\partial \Vg}{\partial h} \, .
\label{meridian}
\eeq
By using this equation we can eliminate one of the variables $x,y$ from
$\Vt$, and the result is a potential depending on $\frakm$.
By extremizing this potential, 
we recover a polynomial in $e^{\frakm}$ and $e^{\frakl}$, the A-polynomial
\cite{CooperApolynomial} of the mapping torus. 

We also analyze subleading contributions to our trace and find that it
reproduces the Reidemeister(-Ray-Singer) torsion of the 3-manifold,
which is the 1-loop contribution in the Chern-Simons theory \cite{Witten:1988hf}.

We stress we do not need to invoke any conjectures in the logic given 
above\footnote{See the comment in footnote \ref{footnote1}, however.};
for example, we do not need to assume AGT conjecture nor 
the equivalence of quantum 
Liouville theory and quantum \Teichmuller theory.
The partition function of our 3d $\scN=2$ theory is computed exactly by 
localization, and we can analyze its
semiclassical limit without any ambiguity.
The only fact we need is that the partition function
can be expressed as an expectation value of an operator
in a certain well-defined Hilbert space, which is our case is the
quantum \Teichmuller space. This naturally realizes the change of
variables\footnote{This refers to the change between Fock coordinates
and Fenchel-Nielsen coordinates.} needed for direct comparison with hyperbolic volume.

%--------------------------------------------------
\section{Semiclassical Limit of the Trace}\label{sec.semiclassical}

We are going to describe the semiclassical limit of our trace
in quantum \Teichmuller  theory.

The Hilbert space of the quantum \Teichmuller theory for the once-punctured
torus (see Sec.\ 4.3 of \cite{Terashima:2011qi})
is spanned by
$\sfx, \sfy$ and $\sfz$, whose non-trivial commutation relations are given by
\beq
[\sfx,\sfy]=[\sfy,\sfz]=[\sfz,\sfx]=-4 \pi i b^2\, ,
\label{CCR}
\eeq
or equivalently
\beq
\sfX \sfY=q^{-4}\sfY \sfX, \quad \sfY \sfZ=q^{-4}\sfZ \sfY, \quad
\sfZ\sfX=q^{-4}\sfX \sfZ\, ,
\label{XYZalg}
\eeq
where in the following capitalized variables represent exponentiation, 
\beq
\sfX=e^{\sfx}, \quad \sfY=e^{\sfy}, \quad \sfZ=e^{\sfz}\, ,
\eeq
and $q:=e^{i\pi b^2}.$
This algebra has a central element, i.e. a constant, corresponding to
the size of the hole
\beq
h:=\sfx+\sfy+\sfz\, .
\eeq
There are 2 remaining variables $\sfx$ and $\sfy$, and we can choose a basis $|x \rangle$ such that
\beq
\sfx|x\rangle=x |x\rangle , \quad \sfy |x\rangle =4\pi i
b^2\frac{\partial}{\partial x} |x \rangle\, .
\eeq
This is a complete set
\beq
\int dx |x \rangle \langle x|=1\, .
\eeq
Similarly, we can choose a different basis $|y\rangle$. This again 
spans a complete set.

The mapping class group $SL(2,\bZ)$ acts on this Hilbert space.
For this purpose it is useful to choose the generators of $SL(2,\bZ)$:
\beq
L=\left( \begin{array}{cc}
  1 & 1\\
  0 & 1
\end{array}
\right) ,\quad
R=\left( \begin{array}{cc}
  1 & 0\\
  1 & 1
\end{array}
\right)\, .
\eeq

As explained in \cite{Terashima:2011qi}, the action of $L$
has the effect
\begin{equation}
\begin{split}
\sfX & \to \mathsf{L}^{-1}\, \mathsf{X}\, \mathsf{L}
 =(1+q\sfX^{-1})^{-1} (1+q^3 \sfX^{-1})^{-1} \sfZ\, , \quad
 \\ 
\sfY &\to \mathsf{L}^{-1}\, \mathsf{Y}\, \mathsf{L}=(1+q\sfX)(1+q^3 \sfX)
\sfY \, , \quad \\
\sfZ & \to \mathsf{L}^{-1}\, \mathsf{Z}\, \mathsf{L}=\sfX^{-1} \, , 
\label{Loperator}
\end{split}
\end{equation}
This preserves the commutation relations given in \eqref{CCR}.
The operator $\sfL$, representing an action of $L$ in $SL(2,\bZ)$, can
be written as a product of two operators\footnote{Essentially the same
decomposition can be found in \cite{FockGoncharovEnsembles}.}
\beq
\sfL=\sfL(\sfX,\sfZ)=f_L(e^{\sfx+\sfz}) g_L(\sfX)=f_L(q^{-2} \sfX\sfZ)\,
g_L(\sfX)\, ,
\eeq
where $f_L(e^{x+z})$ and $g_L(X)$ are given by
\beq
f_L(e^{x+z})=\exp \left[\frac{1}{8\pi i b^2} (x+z)^2\right], \quad
g_L(X)=e_b\left(\frac{x}{2\pi b}\right)^{-1}\, ,
\eeq
where $e_b(z)$ is the quantum dilogarithm function defined in Appendix
\ref{sec.dilog}. 
These functions satisfy (see \eqref{eb2shift})
\beq
f_L(q^4X)= X f_L(X), \quad g_L(q^4 X)=(1+qX)(1+q^3 X) g_L(X)\, .
\eeq
Note that in the convention here the operator $f_L(q^{-2} \sfX \sfZ)$ acts first, then
$g_L(\sfX)$. 
Conjugation by $f_L(q^{-2} \sfX\sfZ)$ acts as
\beq
\sfX\to q^{-4} \sfX^2 \sfZ, \quad \sfY\to \sfY, \quad \sfZ \to
\sfX^{-1}\, ,
\eeq
whereas the conjugation by $g_L(\sfX)$ as
\beq
\sfX\to \sfX, \quad \sfY\to (1+q\sfX)(1+q^3 \sfX)\sfY, \quad 
\sfZ\to (1+q^{-1} \sfX)^{-1} (1+q^{-3} \sfX)^{-1} \sfZ\, ,
\eeq
and we can verify that the composition of the two
gives the desired result \eqref{Loperator}.

Similarly, $\sfR$ is given by\footnote{For this purpose it is useful to note that the commutation
relations
are invariant under the simultaneous exchange of $\sfX, \sfY$ and 
$q, q^{-1}$.}
\beq
\sfR=\sfR(\sfY,\sfZ)=f_R(e^{\sfy+\sfz}) g_R(\sfY)=f_R(q^2\sfY \sfZ)
g_R(\sfY)\, ,
\label{Ropdef}
\eeq
where $f_R$ and $g_R$ are given by
\beq
f_R(e^{y+z})=\exp \left[-\frac{1}{8\pi i b^2} (y+z)^2\right],\quad
g_R(Y)=e_b\left(- \frac{y}{2\pi b}\right)\, .
\eeq

In general an element of the mapping class group is obtained as a
product of these two operators. 
Namely, when we have $\varphi=L^{n_1}R^{m_1} L^{n_2} R^{m_2} \cdots$, we
have\footnote{Let us comment on the ordering of operators. As an example,
$\varphi=LR$ is represented by
\beq
\varphi=\sfL(\sfX,\sfZ)\, \sfR(\sfL^{-1} \sfY \sfL, \sfL^{-1}\sfZ\sfL)\,
.
\eeq
Note that in this expression the argument inside $\sfR$ is conjugated by
an action of $\sfL$. 
This is because the second operator $\sfR$ should act on transformed
variables $\sfL^{-1} \sfY \sfL$ and $\sfL^{-1} \sfZ \sfL$, rather
than the original variables.
However, this expression is equivalent to
\beq
\varphi=\sfR(\sfY,\sfZ)\sfL(\sfX,\sfZ)\, .
\eeq
This means that in the Schr\"odiner representation all we need to do is
to multiply the ket vector from the left (meaning we read the operators
from right to left), while keeping the arguments for the operators
to be the same variables $\sfX, \sfY$ and $\sfZ$. }
\beq
\varphi
|~\rangle=\cdots \sfR(\sfY,\sfZ)^{m_2}\sfL(\sfX,\sfZ)^{n_2}\sfR(\sfY,\sfZ)^{m_1}\sfL(\sfX,\sfZ)^{n_1}
|~\rangle\, .
\eeq

%----------------------------------------------------------
\subsection{Examples: $\varphi=LR$}
Let us now discuss concrete examples.
For $\varphi=LR$, the mapping torus is given 
by the complement of the figure eight knot.
The trace is given by
\begin{eqnarray*}
\textrm{Tr}(\varphi)=\int\! dx\, \langle x| f_R(h-\sfx)g_R(\sfy)
 f_L(h-\sfy) g_L(\sfx) |x \rangle\, .
\end{eqnarray*}
By inserting a complete set as in
\begin{equation*}
\sfR \low{y} \sfL \high{x} \, ,
\end{equation*}
this is computed to be
\begin{eqnarray*}
\textrm{Tr}(\varphi)&=&\int \!dx dy\, \langle x| f_R(h-\sfx)g_R(\sfy)|y\rangle \langle
y| f_L(h-\sfy) g_L(\sfx) |x \rangle \\
&=&\int\! dx dy \, f_R(h-x) g_R(y) f_L(h-y) g_L(x) \, .
\end{eqnarray*}
Let us study the classical limit $b\to 0$ of this quantity. 
In this limit, the functions $g_L(x)$ and $g_R(y)$ simplifies to [see \eqref{Liclassical}]
\beq
g_L(x)\to \exp\left[ -\frac{1}{2\pi i b^2} \textrm{Li}_2 (-e^x) \right], \quad
g_R(x)\to \exp\left[ \frac{1}{2\pi i b^2} \textrm{Li}_2 (-e^{-x})
\right]\, ,
\eeq
and similarly for $f_R(x)$ and $g_R(x)$.
Therefore, the classical limit is given by
\beq
\textrm{Tr}(\varphi)\to \int dx dy \exp \left[\frac{1}{2\pi i
b^2}\Vt(x,y;h)\right]\, ,
\label{Trclassicaleg}
\eeq
where
\beq
\Vt(x,y;h)=-\Li_2(-e^{x})-\frac{1}{4}(h-x)^2+
\Li_2(-e^{-y})+\frac{1}{4}(h-y)^2\, .
\label{VLR}
\eeq
Note that the quadratic part of $h$ cancel out in this expression. As we
will see, this is a generic feature of the potential in the example of
the once-punctured torus bundles
discussed in this paper.

Let temporarily set $h=0$. 
By extremizing $\Vt\big|_{h=0}$, we have 
\begin{align}
\frac{1}{1+e^x}=e^{-x/2}, \quad
\frac{1}{1+e^{-y}} =e^{+y/2} \, ,
\label{LRextremize}
\end{align}
and the potential is maximized at 
$e^{x}=e^y=\frac{-1+i\sqrt{3}}{2}$. The value of the potential there 
is given by $\Li_2(-e^{2\pi i/3})-\Li_2(-e^{-2\pi i/3})$, whose imaginary
(real) part gives  the hyperbolic volume (Chern-Simons invariant) of the figure eight knot complement.
\beq
\textrm{Vol}(4_1)=2.02988..., \quad \textrm{CS}(4_1)=0 \,  .
\eeq
This is not a coincidence,
and we will present a general proof of this statement 
in Secs.\ 3.3 and 3.4. The discussion below serves as an illustrative example.

\begin{figure}[htbp]
\centering{\includegraphics[scale=0.6]{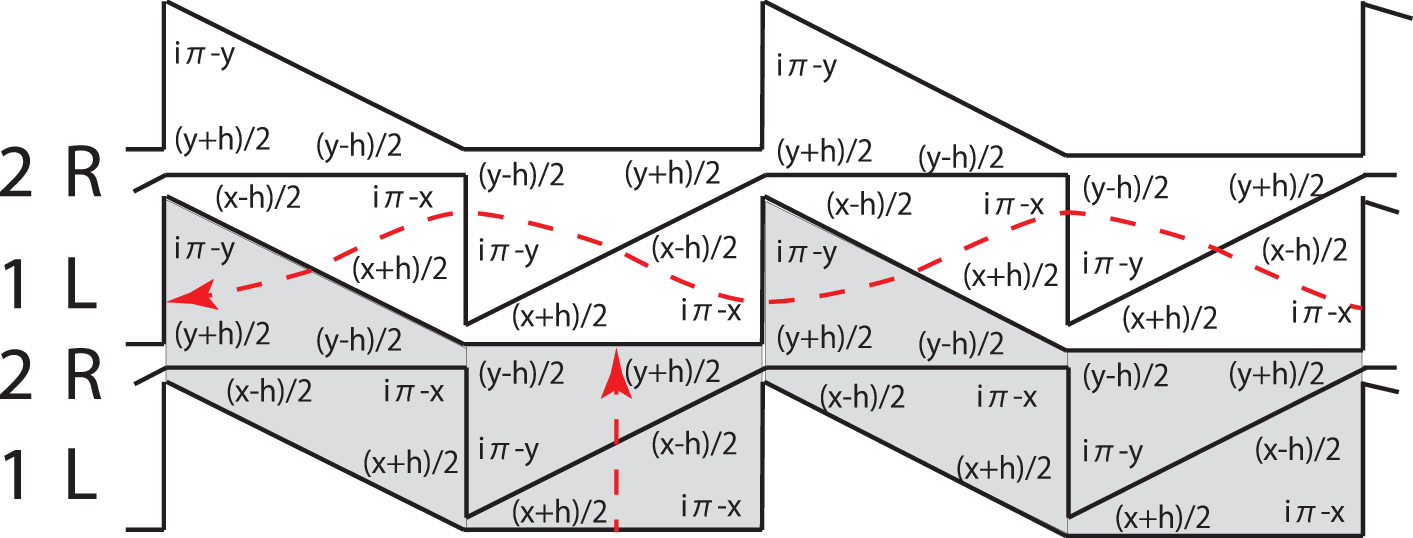}}
\caption{The boundary torus for canonical triangulation of the figure
 eight knot complement. The gray region represents the fundamental
 region
of the torus, and each of the two strips (consisting of four
 triangles)
represents the boundary of a tetrahedron. The modulus of the tetrahedra
 are parametrized by two complex numbers $x$ and $y.$
The horizontal (vertical) dotted arrow represents the longitude
 (meridian) of the torus.}
\label{boundarytorusLR}
\end{figure}

Let us choose the so-called canonical triangulation \cite{FloydHatcher}
of the figure eight knot complement.
In this triangulation we have
two ideal tetrahedra, and in the parametrization of \cite{Gueritaud}
their moduli are
parametrized by two parameters $x$ and $y$ (see Figure
\ref{boundarytorusLR}, with $h=0$).
The consistency conditions for $x,y$ are given by
 \begin{align}
e^{-x/2}=\frac{1}{1-e^{-(i\pi-x)}}, \quad
e^{-y/2}=\frac{1}{1-e^{-(i\pi-y)}} \, , 
\end{align} 
which coincides with \eqref{LRextremize}. 
This shows that $\Vt$ coincides with the $\Vg$ (up to constant terms).

Let us next discuss the $h$ dependence of the potential. Extremizing the
potential, we have
\begin{align}
\frac{1}{1+e^{x}}=e^{-(x-h)/2}, \quad
\frac{1}{1+e^{-y}} =e^{(y-h)/2} \, ,
\label{LRextremizeh}
\end{align}
This is reproduced from the parametrization as in Figure
\ref{boundarytorusLR}.
This 1-parameter deformation preserves the gluing conditions around the
edges of tetrahedra, 
but the longitude parameter is deformed to be \footnote{The complete
hyperbolic metric corresponds to $\frakl=i\pi$.}
\begin{align}
\begin{split}
\frakl &=-(x-h)/2+(i\pi-y)+(x+h)/2-(i\pi-y)+i\pi
=h+i\pi
\, .
\end{split}
\end{align}
This is the 1-parameter deformation of the hyperbolic
structure considered in \cite{NeumannZagier}.
The meridian parameter is given by
\begin{align}
\frakm &=(x+h)/2-(y+h)/2=(x-y)/2=\dfrac{\partial \Vt(x,y;h)}{\partial h}=\frac{\partial \Vt(x,y;\frakl)}{\partial \frakl}
\, .
\label{tmp31}
\end{align}
Note that this is independent of $h$.

Eliminating the variable $y$ using \eqref{tmp31},
the potential becomes
\beq
\Vt(x;\frakm)= -\Li_2(-e^{x})+\Li_2(-e^{-(x-2\frakm)})
-\frac{1}{4}(x^2-(x-2\frakm)^2) +(\frakl-i\pi) \frakm\, .
\eeq
By extremizing this potential with respect to $x$ and $\frakm$
and eliminating the variable $x$,
we have a polynomial equation with respect to
$L:=e^{\frakl}$
and $M:=e^{\frakm}$:
\beq
A({L}, {M})= 1+L(-M^{-2}+M^{-1}+2+M-M^2)+L^2 \, .
\eeq
This coincides with the known expression for the A-polynomial of the
figure eight knot complement.

We can refine our analysis by going to next order, by evaluating
Gaussian integral of the potential $\Vt(x;\frakm)$ around the saddle
point. Explicit computation verifies that this is given by $\sqrt{T(\frakm)}$,
where $T(\frakm)$ is given by
\beq
\frac{1}{\sqrt{(M^2-3M+1) (M^2 + M + 1)}/M}
=\frac{1}{\sqrt{(2\cosh
\frakm-3)(2\cosh \frakm+1)}} \, .
\eeq
This coincides (up to multiplication by an overall constant) with the Reidemeister torsion for the figure eight knot
complement \cite{Porti}.

\subsection{Example: $\varphi=L^2 R$}

Let us next discuss the example of $\varphi=L^2 R$. 
By inserting a complete set as in
\begin{equation*}
\sfR \high{y_1} \sfL\low{x_2, u_2} \sfL \low{x_3} \, ,
\end{equation*}
The trace is computed to be
\begin{align*}
\textrm{Tr}(\varphi)&=\! \int dy_1 dx_2 du_2 dx_3 \, \langle x_3|
 f_R(h-\sfx) g_R(\sfy) |y_1\rangle \langle y_1|
f_L(h-\sfy) g_L(\sfx) |x_2\rangle \\
& \quad \quad \quad \quad \quad \times \langle x_2| u_2\rangle \langle u_2| f_L(h-\sfy) g_L(\sfx)
|x_3\rangle \\
& =\!\int dy_1 dx_2 du_2 dx_3 \, f_R(h-x_3) g_R(y_1) 
f_L(h-y_1) g_L(x_2) f_L(h-u_2) g_L(x_3)
\\
&
 ~~~~~~~~~~~~~~\times
\exp\left[ \frac{1}{4\pi i b^2}
(x_3 y_1-y_1 x_2 +x_2 u_2-u_2 x_3)\right]
\, .
\end{align*}
The classical limit is given by
\beq
\textrm{Tr}(\varphi)\to \int dy_1 dx_2 du_2 dx_3\exp\left[\frac{1}{2\pi
i b^2} \Vt(y_1, x_2, u_2, x_3;h)\right]\, ,
\eeq
where
\begin{equation}
\begin{split}
\Vt(y_1, x_2, u_2, x_3;h)&=
 -\frac{1}{4}(h-x_3)^2+ \Li_2(-e^{-y_1}) 
+\frac{1}{4}(h-y_1)^2-\Li_2(-e^{x_2}) 
\\
&+\frac{1}{4}(h-u_2)^2-\Li_2(-e^{x_3})
+\frac{1}{2}(x_3 y_1-y_1 x_2 +x_2 u_2-u_2 x_3)\, .
\end{split}
\end{equation}
The resulting expression is quadratic in $u_2$, so we can easily integrate out $u_2$, giving
\begin{equation}
\begin{split}
\Vt(y_1,x_2, x_3;h)&=
-\Li_2(-e^{x_3}) -\Li_2(-e^{x_2})  + \Li_2(-e^{-y_1}) 
 -\frac{1}{4}x_3^2
+\frac{1}{4}y_1^2
\\
&-\frac{1}{4}(x_3-x_2)^2
+\frac{1}{2}(x_3 y_1-y_1 x_2)+\frac{1}{2}h(x_2-y_1) \, .
\end{split}
\end{equation}
Note again that this has a linear dependence with respect to $h$.

When we take $h=0$, the potential is maximized by
\begin{equation}
e^{x_3}=\frac{-3 + i \sqrt{7}}{8}, \quad e^{x_2}=e^{y_1}=\frac{-1 + i
 \sqrt{7}}{4} \, ,
\end{equation}
and the imaginary part of the extremal value of the potential
coincides with the volume
of the mapping torus, which is known to be $2.66674...$\, .
The real part, divided by a factor $2\pi^2$ (see \eqref{mainclaim}), gives 
the Chern-Simons invariant $-0.02083$.

From the $h$-linear part, the meridian variable is defined by
\beq
\frakm=\frac{\partial \Vt(y_1, x_2, x_3;h)}{\partial
h}=\frac{1}{2}(x_2-y_1)\, ,
\label{frakmL^2R}
\eeq
and procedures similar to the previous example 
give
\beq
L^2 M + L (- 1/M +2+ 2 M - M^2) +1=0 \, ,
\eeq
which is the known expression for the A-polynomial.
The Gaussian integral around the saddle point is computed to be
$\sqrt{T(\frakm)}$
with 
\beq
T(\frakm)=\frac{1}{\sqrt{1 + M (1 + M) (-2 -3 M+ M^2)}/M}
=\frac{1}{\sqrt{4 \cosh \frakm^2-4\cosh \frakm-7 }}\, .
\eeq
This again coincides with the Reidemeister torsion \cite{Porti}.

\subsection{General $\varphi$: Computation of the Trace}\label{subsec.tracegeneral}

Let us finally discuss the case of a general element $\varphi$ in the mapping class group. 

The element $\varphi$ is represented as a word of $L$ and $R$, and its trace is
again computed by inserting a complete basis of states between $L$'s and $R$'s.
Depending on the four possibilities, we are going to insert complete
sets as in
\begin{equation*}
\cdots \sfL \low{x_k,u_k} \sfL \cdots, \quad \cdots \sfR \high{y_k,v_k}
\sfR \cdots, \quad
\cdots \sfL \low{x_k} \sfR \cdots, \quad \cdots \sfR \high{y_k} \sfL
\cdots
\, .
\end{equation*}
For example, for $\varphi=\sfL \sfL \sfL \sfR \sfR \sfL \sfR \sfR$, we have
\begin{equation*}
 \sfL \low{x_1,u_1} \sfL\low{x_2, u_2} \sfL\low{x_3} \sfR \high{y_4, v_4} \sfR
  \high{y_5} \sfL \low{x_6} \sfR \high{y_7, v_7} \sfR \high{y_8} \, .
\end{equation*}
This gives a potential $\Vt(x,y,u,v;h)$, which after integrating out $u$
and $v$ reduces to the potential $\Vt(x,y;h)$.
In the following we will determine the $x_k$ and $y_k$ dependence of the 
potential $\Vt(x,y;h)$. For this purpose we discuss $2^3=8$
possibilities separately.
The symbol $\equiv$ means equality up to 
terms independent of $x_k$ or $y_k$.
 
In the first four cases, we concentrate on the $x_k$ dependence of the potential.
\begin{enumerate}
\item Suppose we have $\cdots \sfL \sfL \sfL \cdots$. We could then insert
a complete set as in
\begin{equation*}
\cdots \sfL \low{x_{k-1}, u_{k-1}} \sfL \low{x_k, u_k} \sfL \low{x_{k+1}} \cdots,
\end{equation*}
and then we have
\begin{align*}
V_{\rm trace}(x,y,u,v;h)\equiv 
&-\Li_2(-e^{x_k})+\frac{1}{4}\left( (h-u_{k-1})^2+(h-u_k)^2 \right) \\
&\quad \quad+\frac{1}{2}(x_{k-1} u_{k-1}-u_{k-1}x_k+x_k u_k -u_k
 x_{k+1}) \, .
\end{align*}
We can trivially integrate out $u_{k-1}$ and $u_k$,
and the potential becomes
\beq
V_{\rm trace}(x,y;h)\equiv -\Li_2(-e^{x_k})+\frac{1}{4}(-2 x_k^2)
 +\frac{1}{2}x_k(x_{k-1}+x_{k+1}) \, .
\eeq

\item 
For 
$\cdots \sfL \low{x_{k-1}, u_{k-1}} \sfL \low{x_k} \sfR \high{y_{k+1}} \cdots$,
we have
\beq
V_{\rm trace}(x,y;h)\equiv -\Li_2(-e^{x_k})+\frac{1}{4}(-2 x_k^2)
 +\frac{1}{2} x_k(x_{k-1}+y_{k+1}) \, .
\eeq
Note that this is the same as the previous answer except that $x_{k+1}$
      is replaced by $y_{k+1}$.
The same remark applies to all the remaining cases, and each makes a pair with another.

\item 
The analysis is similar for other cases. For 
$\cdots \sfR \high{y_{k-1}} \sfL \low{x_k, u_k} \sfL \low{x_{k+1}} \cdots$,
we have
\beq
V_{\rm trace}(x,y;h)\equiv -\Li_2(-e^{x_k})+\frac{1}{4}(- x_k^2)
 +\frac{1}{2} x_k (x_{k+1}-y_{k-1})+\frac{h}{2} x_k \, .
\eeq

\item 
For $\cdots \sfR \high{y_{k-1}} \sfL \low{x_k} \sfR \low{y_{k+1}} \cdots$,
we have
\beq
V_{\rm trace}(x,y;h)\equiv -\Li_2(-e^{x_k})+\frac{1}{4}(-x_k^2)
 +\frac{1}{2} x_k(y_{k+1}-y_{k-1})+\frac{h}{2} x_k \, .
\eeq

\end{enumerate}

\noindent
The $y_k$ dependence of the potential in the remaining four
cases is determined similarly.

\begin{enumerate}
\setcounter{enumi}{4}
\item 
For $\cdots \sfR \high{y_{k-1}, v_{k-1}} \sfR \high{y_k, v_k} \sfR
      \high{y_{k+1}} \cdots$, we have
\beq
V_{\rm trace}(x,y;h)\equiv \Li_2(-e^{-y_k})+\frac{1}{4}(2 y_k^2)
 +\frac{1}{2}y_k(-y_{k-1}-y_{k+1}) \, .
\eeq

\item 
For $\cdots \sfR \high{y_{k-1}, v_{k-1}} \sfR \high{y_k} \sfL \low{x_{k+1}} \cdots$,
we have
\beq
V_{\rm trace}(x,y;h)\equiv \Li_2(-e^{-y_k})+\frac{1}{4}(2 y_k^2)
 +\frac{1}{2}y_k(-y_{k-1}-x_{k+1}) \, .
\eeq

\item 
For $\cdots \sfL \low{x_{k-1}} \sfR \high{y_k, v_k} \sfR \high{y_{k+1}} \cdots$,
we have
\beq
V_{\rm trace}(x,y;h)\equiv \Li_2(-e^{-y_k})+\frac{1}{4}y_k^2
 +\frac{1}{2}y_k(x_{k-1} - y_{k+1})+\frac{h}{2} (-y_k) \, .
\eeq

\item 
For $\cdots \sfL \low{x_{k-1}} \sfR \high{y_k} \sfL \high{x_{k+1}}
      \cdots$,
we have
\beq
V_{\rm trace}(x,y;h)\equiv \Li_2(-e^{-y_k})+\frac{1}{4}y_k^2
 +\frac{1}{2}y_k(x_{k-1}-x_{k+1})+\frac{h}{2}(-y_k)  \, .
\eeq

\end{enumerate}

We have now determined the function $\Vt(x,y;h)$, since we know the
dependence with respect to all the variables.
In particular, we can extract the $h$-dependent part of the potential.
For 
\begin{equation*}
\sfL^{i_1} \sfR^{j_1} \sfL^{i_2} \sfR^{j_2}   \cdots \sfL^{i_p}
 \sfR^{j_p} \, ,
\end{equation*}
we can insert a complete set as in 
\begin{align*}
& \sfL\low{x_1}\sfL \low{x_2} \cdots\low{x_{i_1-1}} \sfL
\low{x_{i_1}} 
\sfR \high{y_{i_1+1}} \sfR \high{y_{i_1+2}} \cdots \high{y_{i_1+j_1-1}} \sfR
\high{y_{i_1+j_1}} \\
& \hspace{5cm}\sfL\low{x_{i_1+j_1+1}}\sfL \low{x_{i_1+j_1+2}} \cdots
 \sfL \low{x_{i_1+j_1+i_2-1}} \sfL \low{x_{i_1+j_1+i_2}}\sfR
\cdots \, ,
\end{align*}
(here we did not write $u$ and $v$ variables, which we integrate out
anyway), and the $h$-dependent term of $\Vt(x,y;h)$ is
\begin{equation}
\begin{split}
\Vt\supset
\frac{h}{2}(&x_1+x_{i_1+1}+x_{i_1+i_2+1}+\ldots +x_{i_1+\ldots
i_{p-1}+1}
\\
& -y_1-y_{j_1+1}-y_{j_1+j_2+1}-\ldots -y_{j_1+\ldots j_{p-1}+1}
)\, .
\label{hdependence}
\end{split}
\end{equation}

\subsection{General $\varphi$: Computation of Hyperbolic Volume}

The 3-manifold discussed in this paper is a mapping torus of the
once-punctured torus.
This is a knot complement inside a 3-manifold, where 
the position of the knot corresponds to the puncture of the torus.
There is a standard triangulation of the mapping torus, 
called the canonical triangulation \cite{FloydHatcher}.
This is treated in detail in a beautiful paper by Gueritaud \cite{Gueritaud}.

As explained in \cite{Terashima:2011qi}, Sec.\ 4.4, each $L$ or $R$ in
the decomposition of $\varphi$ corresponds to an ideal tetrahedron.
The decomposition of $\varphi$ into $L$ and $R$ represents how to stack 
these ideal tetrahedra $\Delta_k$ one by one.

\begin{figure}[htbp]
\centering{\includegraphics[scale=0.42]{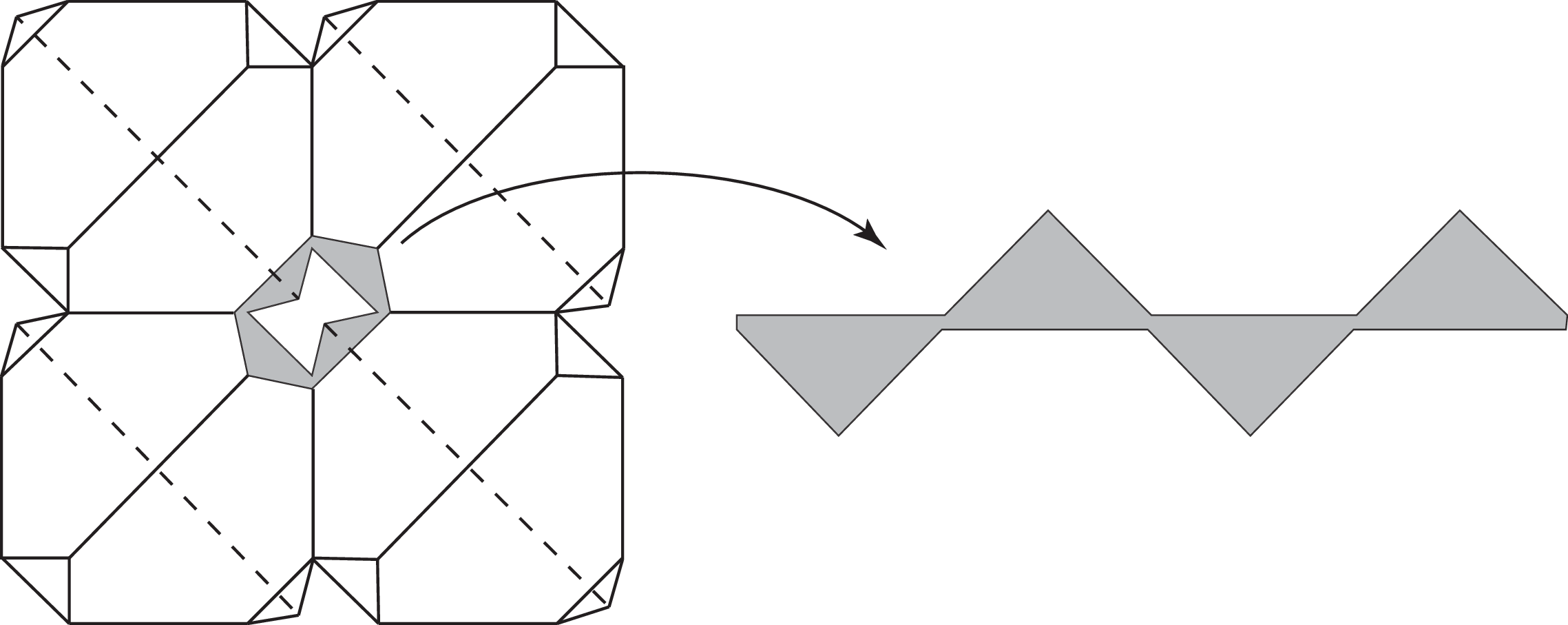}}
\caption{
All the four vertices of an ideal tetrahedron gather around the
 puncture of the torus.
When we cut the tetrahedron around a puncture, the boundary 
is a union of four triangles, colored gray. See \cite{Gueritaud}.}
\label{aroundvertex}
\end{figure}

\begin{figure}[htbp]
\centering{\includegraphics[scale=0.5]{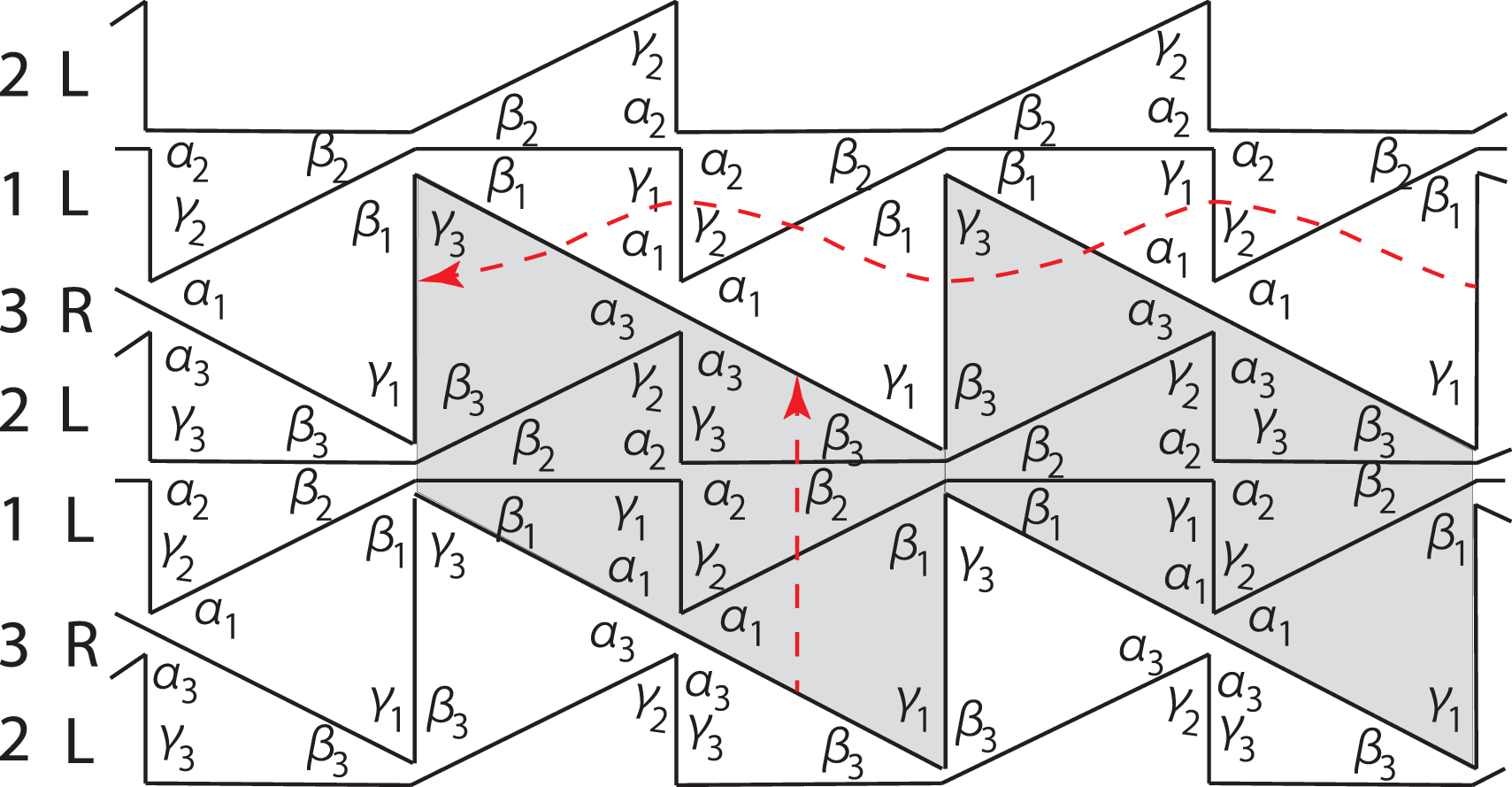}}
\caption{The figure for the boundary torus, for an example 
$\varphi=L^2 R$. The fundamental region of the torus is shown in gray.
We express $\varphi$ as a product of $L$ and $R$, and we layer the four
 triangles of Figure \ref{aroundvertex} in different ways
depending $L$ or $R$. The angles in the figure are not depicted
 correctly,
and angles with the same symbol should really the same.
A more complicated example can be found in Figure 4 of \cite{Gueritaud}.
}
\label{boundarytorus}
\end{figure}

The vertices of the ideal tetrahedron are at the puncture of the torus,
so we need to cut the tetrahedron around the vertices.
Since all the four vertices surround the same vertex, the boundary of the
tetrahedron around the puncture looks like a union of 
four triangles (see Figure \ref{aroundvertex}).
By repeating this for each tetrahedron we have a 
figure for the boundary torus.
See Figure \ref{boundarytorus} for an example.

An ideal tetrahedron has three dihedral angles (see
Appendix \ref{sec.hyperbolic} for this and related materials).
We use a special 
parametrization of the dihedral angles
due to \cite{Gueritaud}. This parametrization is 
given by following the two steps: first, we assign
a variable $w_k$ to each ideal tetrahedron $\De_k$. Second, 
the shape parameters $e^{\alpha_k}, e^{\beta_k}, e^{\gamma_k}$ of 
$\De_k$ are given by the rules in Table \ref{weightrule}, which depends
on whether $L$ or $R$ is assigned to $\Delta_k$ and $\Delta_{k-1}$, respectively.

\begin{table}[htbp]
\caption{Parametrization of the dihedral angles in $\Delta_k$.  
$\alpha_k, \beta_k, \gamma_k$ are (the logarithm of) the shape parameters of
 the tetrahedron $\Delta_k$, see Figure \ref{boundarytorus}.
This is a 1-parameter deformation of the parametrization of
 \cite{Gueritaud}.
We have changed the normalization of $w_k$; $w_k$ here is $2i$ times that in \cite{Gueritaud}.
}
\label{parametrization}
\begin{center}
\scalebox{0.9}{
\begin{tabular}{c||c|c|c|c}
  &   $(\Delta_{k-1}, \Delta_k)=(L,L)$ &  $(\Delta_{k-1}, \Delta_k)=(R, R)$ &  $(\Delta_{k-1}, \Delta_k)=(L,R)$ & $(\Delta_{k-1}, \Delta_k)=(R,L)$  \\
\hline
\hline
$\alpha_k$ & $\dfrac{w_{k-1}+w_{k+1}}{2}$ &$ \dfrac{-w_{k-1}+2 w_k-w_{k-1}}{2}$ &
	    $ \dfrac{w_{k-1}+w_k-w_{k+1}-h}{2}$& $ \dfrac{-w_{k-1}+w_k+w_{k+1}+h}{2}$ \\
\hline
$\beta_k$ & $\dfrac{-w_{k-1}+2 w_k-w_{k+1}}{2}$ & $ \dfrac{w_{k-1}+w_{k+1}}{2}$ &
	     $\dfrac{-w_{k-1}+w_k+w_{k+1}+h}{2}$& $\dfrac{w_{k-1}+w_k-w_{k+1}-h}{2}$ \\
\hline
$\gamma_k$ & $\pi i-w_k$& $ \pi i-w_k$ & $\pi i- w_k$& $ \pi i- w_k$\\
\end{tabular}
} % scalebox
\end{center}
\label{weightrule}
\end{table}

The parametrization chosen here automatically satisfies
the gluing condition for each edge \cite{Gueritaud}.
In the example of Figure
\ref{boundarytorus},
the consistency conditions are  
\begin{equation*}
2(\beta_1+\beta_2+\beta_3)+\gamma_1+\gamma_3=2\pi i,
\quad 2\alpha_2+\gamma_1+\gamma_3=2\pi i ,\quad
2(\alpha_1+\alpha_3+\gamma_2)=2\pi i \, .
\end{equation*}
This is satisfied by the parametrization
\begin{align*}
&\alpha_1=\frac{-w_3+w_1+w_2+h}{2}, \quad \alpha_2=\frac{w_1+w_3}{2}, \quad \alpha_3=\frac{w_2+w_3-w_1-h}{2} \, , \\
&\beta_1=\frac{w_3+w_1-w_2-h}{2},\quad \beta_2=\frac{-w_1+2w_2-w_3}{2}, \quad \beta_3=\frac{-w_2+w_3+w_1+h}{2}\, , \\
&\gamma_1=\pi i-w_1,\quad \gamma_2=\pi i- w_2, \quad \gamma_3=\pi i- w_3 \,  .
\end{align*}

We still have to worry about the consistency conditions 
inside each tetrahedron; the three shape parameters in the 
same ideal tetrahedron must have relations
\begin{align}
 e^{-\alpha_k}=1-e^{\gamma_k}, \quad
 e^{-\beta_k}=\frac{1}{1-e^{-\gamma_k}} \, .
\label{tetraconsistency}
\end{align} 
Note that these two equations are not independent since we have $e^{\alpha_k+\beta_k+\gamma_k}=1$.
Since each of $\alpha_k, \beta_k, \gamma_k$ are written by 
$w_{k-1}, w_k$ and $w_{k+1}$, these are relations among the variables
$w_k$'s. We can construct a potential $\Vg (w)$ whose derivative with respect
to $w_k$ reproduces \eqref{tetraconsistency}. By extremizing the
potential
we find the complete hyperbolic structure on the mapping torus \cite{Rivin}.

In the following we denote $w_k$ by $x_k, y_k$, depending on
whether the corresponding tetrahedron is $L$ or $R$. 
Then the structure equations \eqref{tetraconsistency} will take the
following form, depending whether each of 
$\Delta_{k-1}, \Delta_k,
\Delta_{k+1}$ is of type $L$ or $R$.

\begin{enumerate}
\item  $(\Delta_{k-1}, \Delta_k, \Delta_{k+1})=(L,L,L)$. 

Following the rule in Table \ref{weightrule}, we have
\begin{equation}
\alpha_k=\frac{w_{k-1}+w_{k+1}}{2}, \quad \beta_k=\frac{2 w_k-w_{k-1}-w_{k+1}}{2}, \quad \gamma_k=i\pi- w_k\, .
\end{equation}
We also have 
\begin{equation}
x_j=w_j \quad (j=k-1, k, k+1)\, .
\end{equation}
From these equations the second equation of \eqref{tetraconsistency}
becomes
\begin{equation}
\frac{1}{1+e^{x_k}}=e^{\frac{1}{2}(-2 x_k+x_{k+1}+x_{k-1})} \, .
\end{equation}
This is reproduced from a potential
\begin{equation}
\Vg(x,y;h)\equiv
 -\Li_2(-e^{x_k})+\frac{1}{4}(-2x_k^2)+\frac{1}{2}x_k(x_{k+1}+x_{k-1}) \, ,
\end{equation}
which coincides with the $\Vt(x,y;h)$.
\end{enumerate}

The remaining cases are treated similarly, and we list the structure
equation as well as the potential which reproduces it. Note that the
structure equation itself is determined from $\Delta_{k-1}$ and
$\Delta_k$,
and the $L/R$ type of $\Delta_{k+1}$ changes only the label of the variable.
\begin{enumerate}
\setcounter{enumi}{1}

\item   $(\Delta_{k-1}, \Delta_k, \Delta_{k+1})=(L,L,R)$.
\begin{equation*}
\frac{1}{1+e^{x_k}}=e^{\frac{1}{2}(-2x_k+y_{k+1}+x_{k-1})}, \quad
\Vg\equiv
-\Li_2(-e^{x_k})+\frac{1}{4}(-2x_k^2)+\frac{1}{2}x_k(y_{k+1}+x_{k-1})
\, .
\end{equation*}

\item  $(\Delta_{k-1}, \Delta_k, \Delta_{k+1})=(R,L,L)$.
\begin{equation*}
\frac{1}{1+e^{x_k}}=e^{\frac{1}{2}(-x_k+x_{k+1}-y_{k-1}+h)}, \quad
\Vg\equiv
-\Li_2(-e^{x_k})+\frac{1}{4}(-x_k^2)+\frac{1}{2}x_k(x_{k+1}- y_{k-1})+\frac{h}{2}x_k \, .
\end{equation*}

\item  $(\Delta_{k-1}, \Delta_k, \Delta_{k+1})=(R,L,R)$.
\begin{equation*}
\frac{1}{1+e^{x_k}}=e^{\frac{1}{2}(-x_k+y_{k+1}-y_{k-1}+h)}, \quad
\Vg\equiv
-\Li_2(-e^{x_k})+\frac{1}{4}(-x_k^2)+\frac{1}{2}x_k(y_{k+1}-y_{k-1}) 
+\frac{h}{2}x_k
\, .
\end{equation*}

\item   $(\Delta_{k-1}, \Delta_k, \Delta_{k+1})=(R,R,R)$.
\begin{equation*}
\frac{1}{1+e^{y_k}}=e^{\frac{1}{2}(-y_{k+1}-y_{k-1})}, \quad
\Vg\equiv
-\Li_2(-e^{y_k})+\frac{1}{2}y_k(-y_{k+1}- y_{k-1})
\, .
\end{equation*}

\item   $(\Delta_{k-1}, \Delta_k, \Delta_{k+1})=(R,R,L)$.
\begin{equation*}
\frac{1}{1+e^{y_k}}=e^{\frac{1}{2}(-x_{k+1}-y_{k-1})}, \quad
\Vg\equiv -\Li_2(-e^{y_k})+\frac{1}{2}y_k(-x_{k+1}-y_{k-1})
\, .
\end{equation*}

\item   $(\Delta_{k-1}, \Delta_k, \Delta_{k+1})=(L,R,R)$.
\begin{equation*}
\frac{1}{1+e^{y_k}}=e^{\frac{1}{2}(-y_{k+1}+x_{k-1}-y_k-h)}, \quad
\Vg\equiv -\Li_2(-e^{y_k})+\frac{1}{4}(-y_k^2)+\frac{1}{2} y_k(-y_{k+1}+x_{k-1})
-\frac{h}{2}y_k\, .
\end{equation*}

\item   $(\Delta_{k-1}, \Delta_k, \Delta_{k+1})=(L,R,L)$.
\begin{equation*}
\frac{1}{1+e^{y_k}}=e^{\frac{1}{2}(-x_{k+1}+x_{k-1}-y_k-h)},\quad
\Vg\equiv -\Li_2(-e^{y_k})+\frac{1}{4}(-y_k^2)+\frac{1}{2}y_k(-x_{k+1}+x_{k-1})
-\frac{h}{2}y_k
\, .
\end{equation*}

\end{enumerate}

By comparing these results with the results in Sec.\ 
\ref{subsec.tracegeneral}, we have (up to constant terms)
\beq
\Vt(x,y;h)=\Vg(x,y;h) \, ,
\eeq
where we used an identity of the classical dilogarithm \eqref{Li2sum}.
This is what we wanted to show. This equation demonstrates an
equivalence of the potential {\it before extremization},
and clarifies the geometrical meaning of Fock variables $x_k, y_k$ in
hyperbolic geometry.
Moreover, explicit computations show that, for all the 8 cases,
\beq
\textrm{Im}\left( \Vt(x,y,;h)\Big|_{h=0} \right)=
\sum_k D(e^{\gamma_k})
+ \sum_k \textrm{Re}(w_k)
\, \textrm{Im}\left( \frac{\partial V}{\partial w_k}\right) \, ,
\eeq
where $D(z)$, the Bloch-Wigner function defined in \eqref{BlochWigner},
is the hyperbolic volume of an ideal tetrahedron with modulus $z$.
This shows that the $\Vt$ at the critical point gives precisely the
hyperbolic volume of the 3-manifold.

We can also identify longitude and meridian parameters.
The parameter $h$ is identified with $\frakl$,
and the expression inside the bracket of \eqref{hdependence}
coincides
with a holonomy of a cycle along 
the boundary torus.
For example, in Figure \ref{boundarytorus},
the holonomy along the vertical direction gives
\begin{equation*}
\frakm=\alpha_1-\beta_2-\beta_3=\frac{-w_1+w_3}{2}=\frac{x_1-y_3}{2} \, ,
\end{equation*}
whereas the one along the horizontal direction is
\begin{equation*}
\frakl=-\gamma_3+(\alpha_1+\gamma_2)-\beta_1 + i\pi=h +i\pi \, .
\end{equation*}
The general proof can be given similarly. Note meridian parameter
$\frakl$ is given as an $h$-linear term in the potential of the quantum
\Teichmuller theory, and does not require extra input --- this is in
sharp contrast with the state sum model of \cite{HikamiGeneralized},
where the longitude constraint is put in by hand as a delta function.

\section{Concluding Remarks}\label{sec.conclusion}

In this paper we have demonstrated that the classical limit of the partition
functions of 3d theory gives the hyperbolic volume (and Chern-Simons
invariant)
of the 3-manifold,
in the case of the once-punctured torus. 

It would be interesting to extend the discussion to a more general
Riemann surface, 
and to discuss perturbative corrections to
the volume (see \cite{Dijkgraaf:2009sb,Dimofte:2009yn,Dijkgraaf:2010ur}). 

%%%%%%%%%%%%%%%%%%%%%%%%%%%%%%%%%%%%%%%%%%%%%%%%%%%%%%%%%%%%%%%%%%%%%%%%%%%%%%
\section*{Acknowledgments}
The content of this paper was presented by M.~Y. at the String-Math 2011
conference (University of Pennsylvania, June 2011), and we thank the
audience for feedback.
Y.~T. is supported in part by the Grants-in-Aid for
Scientific Research, JSPS. M.~Y.~ thanks PCTS for its support.
We would like to thank T.~Dimofte, H.~Fuji, D.~Gaiotto, T.~Nishioka, and
Y.~Tachikawa for discussion. 
%%%%%%%%%%%%%%%%%%%%%%%%%%%%%%%%%%%%%%%%%%%%%%%%%%%%%%%%%%%%%%%%%%%%%%%%%%%%%

\appendix

%---------------------------------------------------------------------------\

%%%%%%%%%%%%%%%%%%%%%%%%%%%%%%%%%%%%%%%%%%%%%%%%%%%%%%%%%%%%%%%%%%%%%%%%%%%%%
\section{Quantum Dilogarithm} \label{sec.dilog}

In this appendix we collect formulas for the noncompact quantum
dilogarithm function $e_b(z)$ \cite{FaddeevVolkovAbelian,FaddeevKashaevQuantum,Faddeev95}. 

The function $e_b(z)$ is defined by
\beq
e_b(z)=\exp\left(\frac{1}{4} \int_{-\infty+i0}^{\infty+i0}  \frac{dw}{w}
\frac{e^{-i 2zw} }{\sinh(wb) \sinh(w/b)}\right)\, , 
\label{ebdef}
\eeq
where the integration contour is chosen above the pole $w=0$,
and we require $|\mathrm{Im}\, z| <|\mathrm{Im}\, c_b|$ for convergence
at infinity. 
In the classical limit $b\to 0$, we have
\beq
e_b(z)\to \exp \left(\frac{1}{2 \pi i b^2} 
  \textrm{Li}_{2}(-e^{2\pi b z}) \right)\, ,
\label{Liclassical}
\eeq
where $\Li_2(z)$ is the Euler classical dilogarithm function
defined by
\beq
\textrm{Li}_2(z)=-\int_0^z \frac{\log(1-t)}{t}dt\, . 
\label{Li2}
\eeq
This function satisfies
\beq
\Li_2(-e^x)+\Li_2(-e^{-x})=-\frac{\pi^2}{6}-\frac{1}{2}x^2\, .
\label{Li2sum}
\eeq
After analytic continuation, $e_b(z)$ has a product expression when $\textrm{Im}\, b^2>0$:
\beq
e_b(z)=(e^{2\pi (z+iQ/2)b}; q^2)_{\infty}/ (e^{2\pi (z-iQ/2)b^{-1}};
\bar{q}^2)_{\infty}\, , 
\label{dilogproduct}
\eeq
where $(x;q)_{\infty}$ is defined by
\beq
(x;q)_{\infty}=\prod_{n=1}^{\infty} (1-x q^n) \, ,
\eeq
and $q:=e^{i\pi b^2}, \quad \bar{q}:=e^{-i \pi b^{-2}}$.
This implies 
\begin{equation}
\begin{split}
e_b(z-2ib)&=(1+q^{-1} e^{2\pi b z})(1+q^{-3}e^{2\pi b
z}) \,e_b(z) \, ,\\
e_b(z+2ib)&=(1+q e^{2\pi b z})^{-1}(1+q^3e^{2\pi b
z})^{-1} \,e_b(z) \, .
\end{split}
\label{eb2shift}
\end{equation}

%---------------------------------------------------------------------------\

\section{Rudiments of Hyperbolic Geometry} \label{sec.hyperbolic}

In this section we quickly summarize some basics facts of hyperbolic
geometry needed for the understanding of this paper.
See, for example, \cite{ThurstonReview} for further introduction.

Let $\bH^3$ be a 3d hyperbolic space, namely a upper half plane
\beq
\bR^3_{> 0}=\{(x_1, x_2, y) |\, x_1, x_2\in \bR, y>0 \} \, ,
\eeq
with the
metric
\beq
ds^2=\frac{(dx_1)^2+(dx_2)^2+dy^2}{y^2} \, .
\eeq  
This has a boundary $\partial H^3=\bC\cup \{\infty\}$,
and has isometry $PSL(2,\bC)$. 

A hyperbolic manifold can be written in the form $\bH^3/\Gamma$, where
$\Gamma$ is a torsion free discrete subgroup of $PSL(2,\bC)$.
The examples discussed in this paper are knot complements.
Suppose that we have a 3-manifold $M$, and a knot $K$ inside. The knot
complement of $K$ inside $M$ is a complement of the tubular neighborhood
$N(K)$ of $M$
\beq
M\backslash K:=M\backslash N(K) \, .
\eeq
By construction the boundary of $M\backslash K$ is the boundary of $N(K)$, which is
a torus. The cycle of the torus contractible (noncontractible) in $N(K)$ is called the meridian
(longitude). 

\begin{figure}[htbp]
\centering{\includegraphics[scale=0.3]{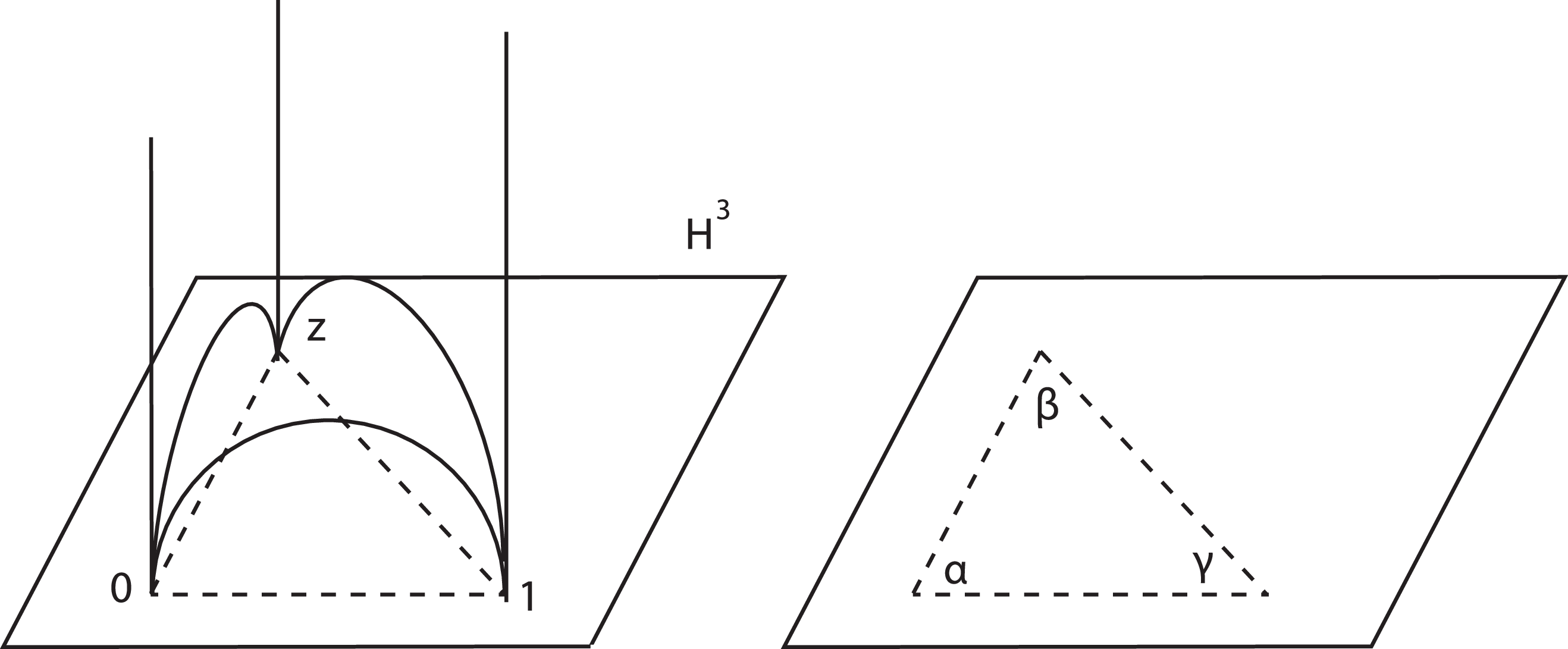}}
\caption{An ideal tetrahedron in $\bH^3$ 
has all the four vertices on
 the boundary of $\bH^3$,
which we can take to be 
$\{0,1,z,\infty\} \in
 \bC\cup \{\infty\}$.}
\label{idealtetrahedron}
\end{figure}

An ideal tetrahedron is a tetrahedron whose all four vertices are on the
boundary of $\bH^3$ (see Figure \ref{idealtetrahedron}). By a suitable M\"obius transformation we can take the vertices to be
at positions $0,1,z$ and infinity. This parameter $z$ is called the
modulus (shape parameter) of the tetrahedron.

A tetrahedron has 6 edges and correspondingly 6 face angles. For an ideal
tetrahedron with modulus
$z$, the two face angles on the opposite side of the tetrahedron are
the same. These are given as the arguments of three complex parameters
\beq
z=e^{\alpha}, \quad \frac{1}{1-z}=e^{\beta}, \quad 1-z^{-1}=e^{\gamma} \, ,
\eeq
satisfying $e^{\alpha+ \beta +\gamma}=1$ and the relations as in \eqref{tetraconsistency}.
The hyperbolic volume of
an ideal tetrahedron with modulus $z$ is given by the Bloch-Wigner
function $D(z)$:
\beq
D(z)=\textrm{Im}\left(\Li_2(z) \right)+\textrm{arg}(1-z) \log|z| \, ,
\label{BlochWigner}
\eeq
which satisfies
\beq
D(z)=D(1-z^{-1})=D\left( \frac{1}{1-z}\right)
=
-D(z^{-1})=-D(1-z)=-D\left( \frac{1}{1-z^{-1}}\right) \, .
\eeq

Let us now glue the tetrahedron to construct 3-manifolds.
There are two types of boundary conditions we need to impose.  
First, we need a gluing condition around an edge or the triangulation:
\beq
\prod_{\rm edge} z_i=1 \, .
\label{bulkgluing}
\eeq
This says that the angles around an edge sum up to $2\pi$.
This condition is already taken care of in the parametrization
of \cite{Gueritaud} explained in the main text.
There are also gluing conditions along the torus boundaries.
This can be written as
\beq
\prod_{\rm meridian} z_i=M, \quad \prod_{\rm longitude} z_i=L^2 \, .
\label{boundarygluing}
\eeq
Here the two parameters $M$ ($L$) are called meridian (longitude)
parameters, and their logarithms are denoted by $\frakm$ ($\frakl$).
In Chern-Simons theory they are the holonomies
along $\alpha$ and $\beta$-cycles.
\beq
\rho(\alpha)=
\left(
\begin{array}{cc}
M^{1/2} & * \\
0 & M^{-1/2}
\end{array}
\right),
\quad
\quad \rho(\beta)=
\left(
\begin{array}{cc}
L & * \\
0 & L^{-1}
\end{array}
\right) \, ,
\eeq
If we take $M=1$, we have a complete hyperbolic metric, and there are
only isolated solutions of \eqref{bulkgluing} and \eqref{boundarygluing}.
The parameter $M$ is a 1-parameter deformation of the hyperbolic
structure \cite{NeumannZagier}.

%---------------------------------------------------------------------------\

%%%%%%%%%%%%%%%%%%%%%%%%%%%%%%%%%%%%%%%%%%%%%%%%%%%%%%%%%%%%%%%%%%%%%%%%%%%%%%%%

\bibliography{AGTknot,AGTadd}
\bibliographystyle{h-physrev5}

%%%%%%%%%%%%%%%%%%%%%%%%%%%%%%%%%%%%%%%%%%%%%%%%%%%%%%%%%%%%%%%%%%%

\end{document}